\documentclass[referee]{raa}            

\usepackage{graphicx,times}             

\graphicspath{{../fig/}}
\usepackage{natbib}
\usepackage{amssymb,amsmath}
\usepackage{multirow}
\usepackage{threeparttable}
\usepackage{subcaption}
\bibpunct{(}{)}{;}{a}{}{,}
\usepackage{caption}
\usepackage[pagebackref=true]{hyperref}

\begin{document}

  \title{Temperature and wind characteristics of Lenghu site for ventilation and structural design of large telescope enclosure}

   \volnopage{Vol.0 (20xx) No.0, 000--000}      
   \setcounter{page}{1}          

   \author{Tao-Ran Li %
      \inst{1,2}, Lu Feng\inst{1,2}, Liang Ge\inst{1,2}, Yong Zhao\inst{1}, Fan Yang\inst{1}, Li-Cai Deng\inst{1,2}, Zhi-Yong Gao\inst{1}, Ying Wu\inst{1} 
   }

   \institute{National Astronomical Observatories, Chinese Academy of Sciences, Beijing 100101, China; {\it litaoran@nao.cas.cn, jacobfeng@nao.cas.cn}\\
        \and
             School of Astronomy and Space Science, University of Chinese Academy of Sciences, Beijing 100049, China;\\
\vs\no
   {\small Received 20xx month day; accepted 20xx month day}}

\abstract{ In recent years, a significant number of observatories and universities have been planning to construct optical and infrared telescopes at the Lenghu site in Qinghai Province due to the site's excellent seeing and clear night sky fraction. Although astronomical performances of the Lenghu site have been reported in detail by numerous papers, there were few reports showing statistics of temperature and wind characteristics in the traditional way required for the design of steel structures of large astronomical telescopes and enclosures, as well as the ventilation and air conditioning systems of these enclosures. This paper aims to present such new statistical data on temperature and wind conditions at the site, which could be helpful to inform and aid in such design decisions at the Lenghu site.
}

   \authorrunning{T.-R. Li et al.}            
   \titlerunning{Temperature and wind characteristics of Lenghu site}  

   \maketitle

%
%
\section{Introduction}           
\label{sect:intro}
The Chinese astronomical community is planning the construction of a range of optical/infrared telescopes with apertures from 1 m to 14.5 m. Identifying a site with good astronomical characteristics is essential to ensure that the telescopes can operate with full potentials. Among the optical/infrared nighttime astronomical sites that have been monitored in recent years, Lenghu has become one of the most popular candidate sites for these telescopes. Several papers (e.g., \citep{deng2021lenghu,2024MNRAS.535.1278L,2024PrA....42..257Y}) have reported on its astronomical performance, such as seeing, cloud coverage, and meteorological parameters, suggesting the site has favorable conditions for nighttime astronomical observations.

The aforementioned papers were primarily focused on reporting the excellence of astronomical performances of the site. However, from an engineering point of view, their statistics are not sufficient or consistent with the "traditional" statistics which are required for designing steel structures (such as the supports and tubes of telescopes and enclosures) and the ventilation/air conditioning system of the enclosure. Therefore, it is necessary to reevaluate certain parameters using different methods rather than solely relying on the site’s previously reported astronomical performances. Three crucial parameters have been identified for further study to assist in the design of telescopes and enclosures: temperature, wind speed, and wind direction. These statistics for platform A and adjacent platforms at the Lenghu site are especially important because these closely packed platforms are already being considered for the placement of telescopes with diameters of 14.5 m, 6.5 m, 3 m, as well as other smaller telescopes.

These parameters were chosen based on the following considerations:

- Temperature: Temperature fluctuations induce thermal expansion and contraction in both the main structure of telescope and the steel structure of enclosure, potentially compromising their mechanical precision and operational stability. Therefore actively control method for the daytime indoor environment needs to be considered especially for large telescopes. During daytime operation, air condition system could help to maintain the dome interior temperature consistent with operation conditions \citep{2010SPIE.7738E..0BV}. During nighttime observation, thermal inhomogeneity inside the dome directly contributes to dome seeing effect \citep{1997SPIE.2871..726Z}, degrading the image quality. 

- Wind speed: Wind speed is a critical factor determining the magnitude of wind loads acting on telescopes and enclosures. The estimation of wind loads is required early in the design stage as it significantly affects the design of enclosure, telescope structure, and control system \citep{2006ApOpt..45.7912M}. Understand wind speed characteristics will be conducive to wind control system design (e.g., vents and louvers) and the natural ventilation frequency. These considerations are essential for maintaining a stable airflow and thermal environment within the enclosure when it is opened for observation.

- Wind direction: Wind direction dictates the layout of the selected site. This factor must be considered to ensure that the placement of telescope minimizes adverse effects from turbulence by topography and other buildings \citep{2022RAA....22d5002L,bely2003design}. Also, real-time variations in wind direction require dynamic adjustments to ventilation strategies, enabling optimization of airflow and maintaining observing environment for the telescope.

This paper describes the temperature, wind speed, and wind direction characteristics at the Lenghu site especially for platform A and C (Lenghu-A and Lengh-C). In contrast to site selection, the study focuses on implications for large-scale telescope and enclosure design, including steel structure, air condition, and ventilation. Meteorological data from Lenghu-A and Lenghu-C were collected, and the analysis focused on hourly temperature and wind speed, sunrise-to-midnight temperature differentials, and time-series variations in wind direction.


\section{Data and Method} \label{sec:2}
\subsection{Lenghu site} \label{sec:2.1}
\begin{figure}[ht!]
  \centering
  \begin{subfigure}{0.45\textwidth}
    \centering
    \includegraphics[height=6cm]{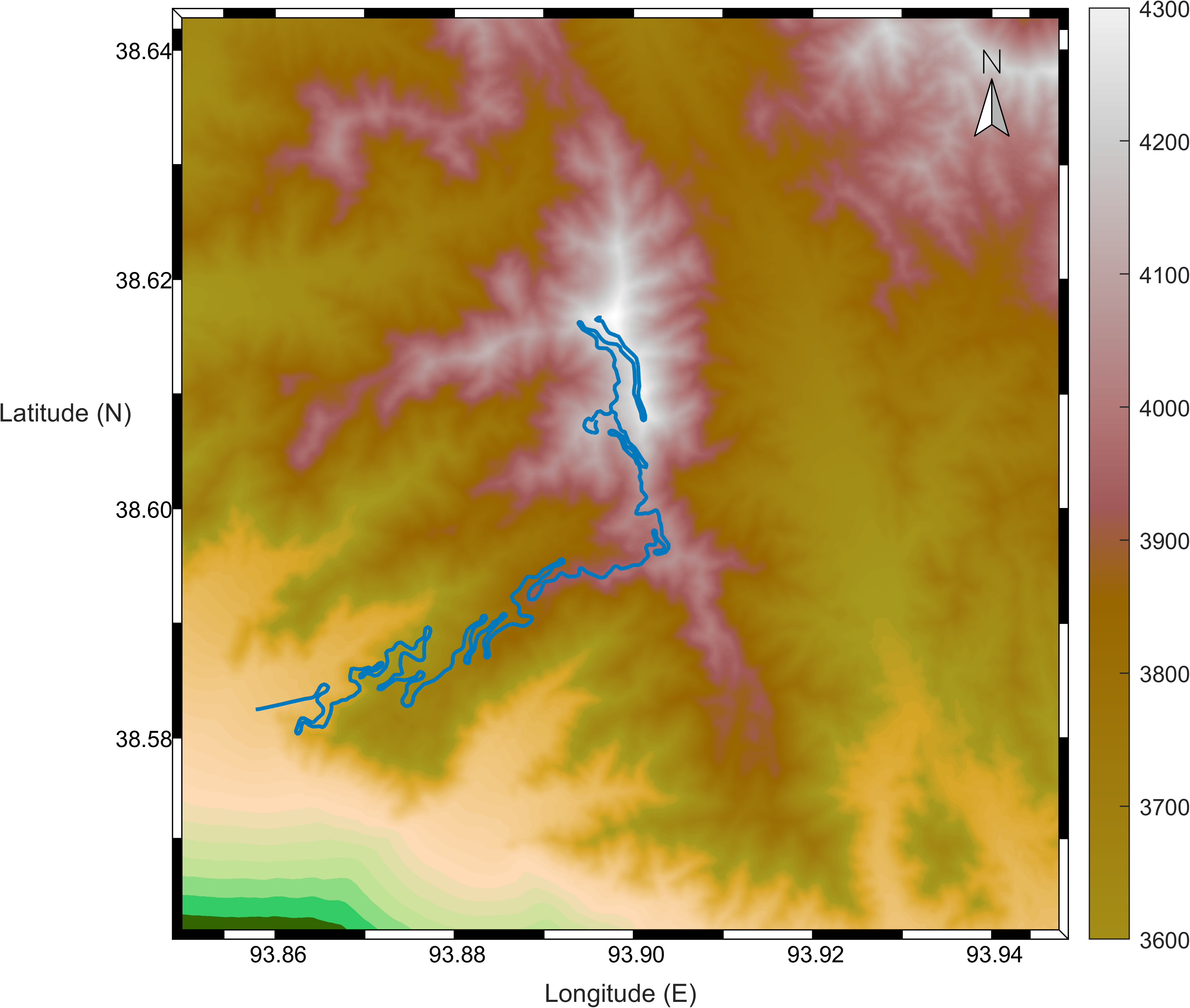}
  \end{subfigure}
  \hspace{1em}
  \begin{subfigure}{0.45\textwidth}
    \includegraphics[height=6cm]{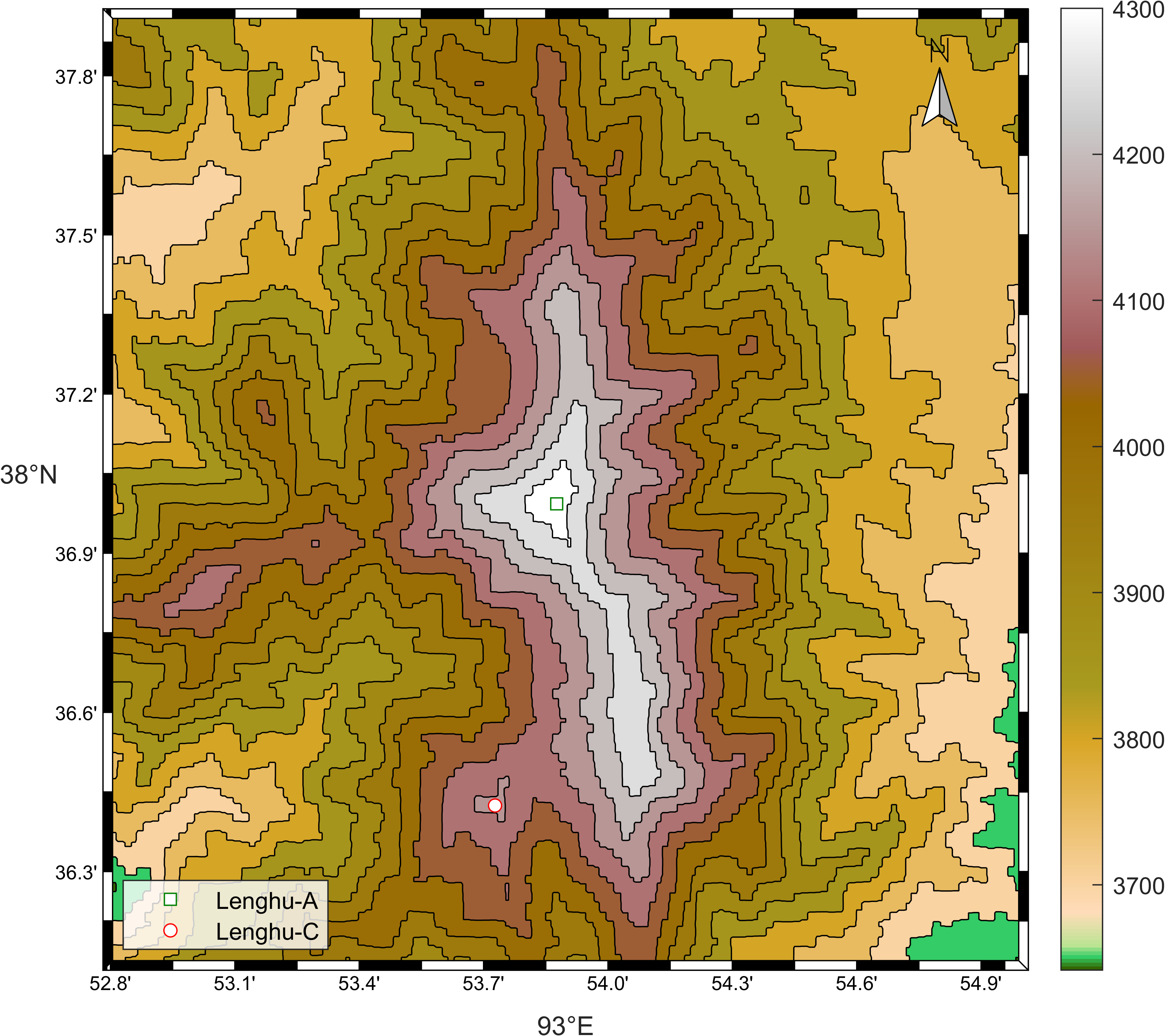}
  \end{subfigure}
  \caption{\textit{Left}: Topographic map of the area of Saishiteng Mountain for Lenghu observatory. The blue curve shows the road from the foot of the mountain to the peak. The road information was extracted from OpenStreetMap.org. \textit{Right}: The contour map of Lenghu-A and Lenghu-C area (marked as green square and red circle, respectively) with a contour interval of 50 m. The digital elevation data was derived from the Copernicus Digital Elevation Model (CopDEM GLO-30, https://doi.org/10.5270/ESA-c5d3d65) and corrected with local topographical survey and mapping data.}
  \label{fig1}
\end{figure}

The Lenghu site comprises multiple platforms distributed across the summits of Saishiteng Mountain in Qinghai Province. It features complex terrain with an average elevation of 4000 m (Figure \ref{fig1}) and exhibits a highland continental climate \citep{2023MNRAS.522.1419Z} characterized by cold, windy conditions and significant diurnal temperature variations, these severe environments impose stringent requirements on the design of telescopes and enclosures. According to on-site survey and mapping, the elevation of Lenghu-A is 4304 m, and that of Lenghu-C is 4163 m.

\subsection{Meteorological station and data} \label{sec:2.2}
\begin{figure}[ht!]
  \centering
  \begin{subfigure}{0.4\textwidth}
    \includegraphics[width=\linewidth]{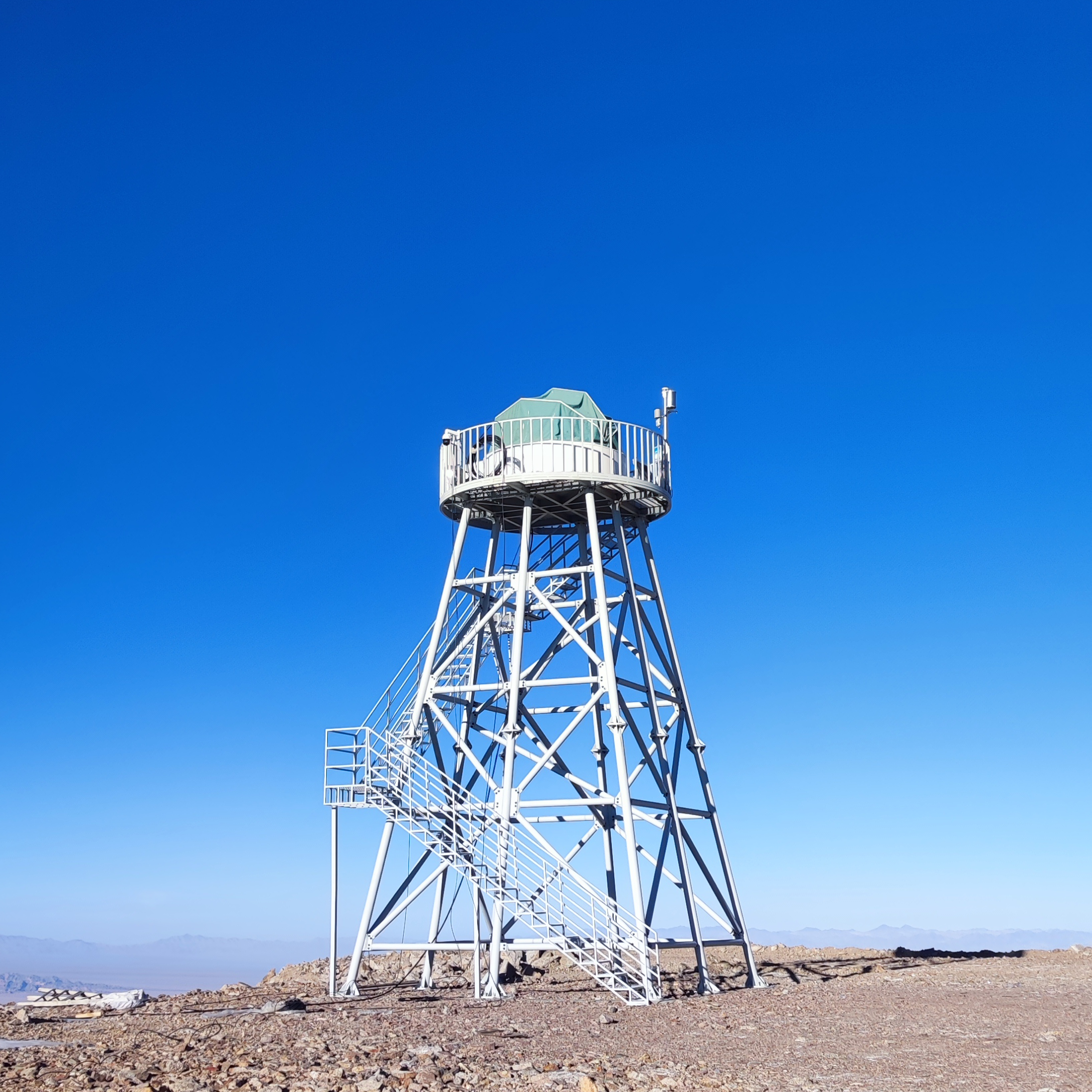}
  \end{subfigure}
  \hspace{1em}
  \begin{subfigure}{0.4\textwidth}
    \includegraphics[width=\linewidth]{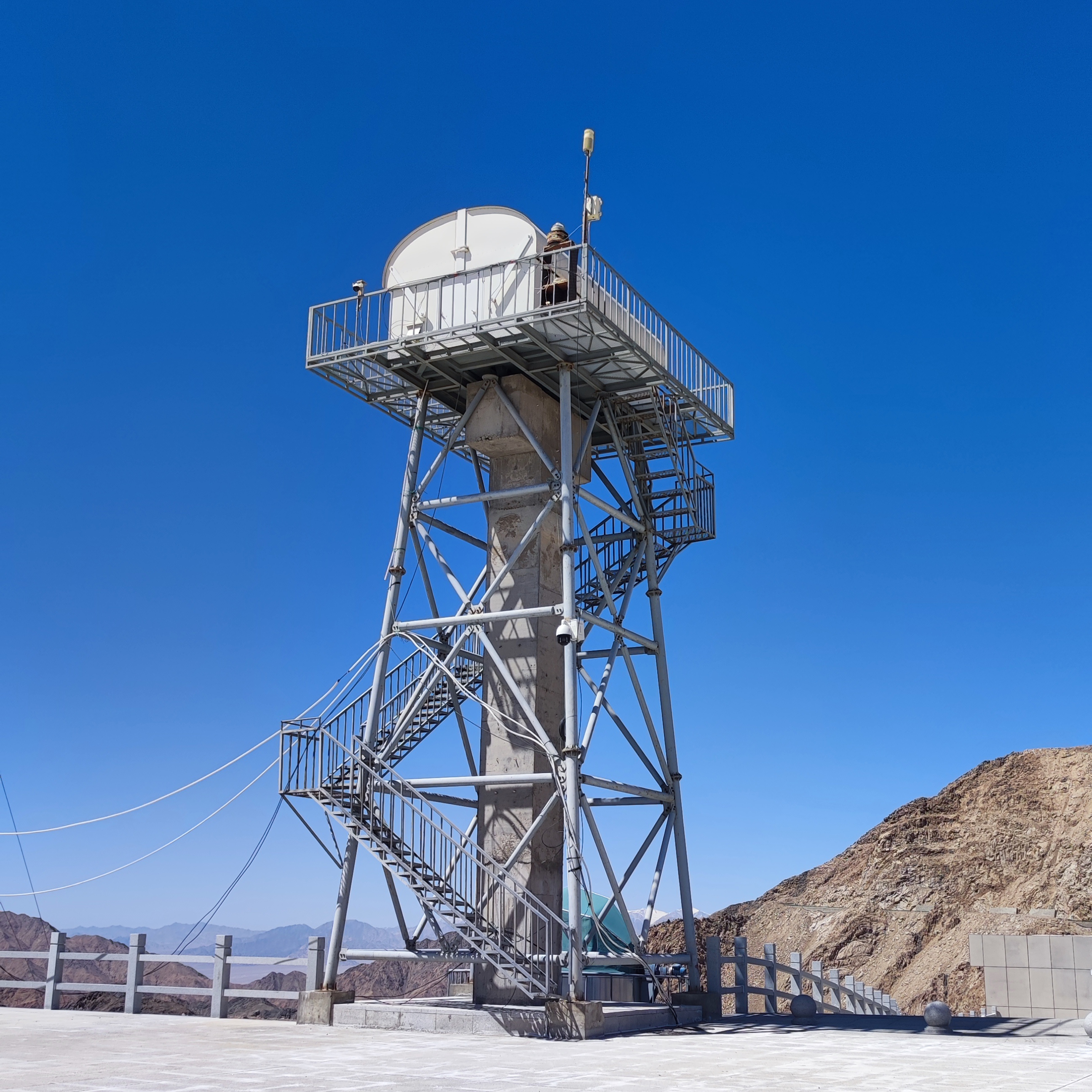}
  \end{subfigure}
  \caption{Meteorological tower at Lenghu-A (left) and Lenghu-C (right)}
  \label{fig2}
\end{figure}

\begin{figure}[ht!]
  \centering
  \includegraphics[width=10cm]{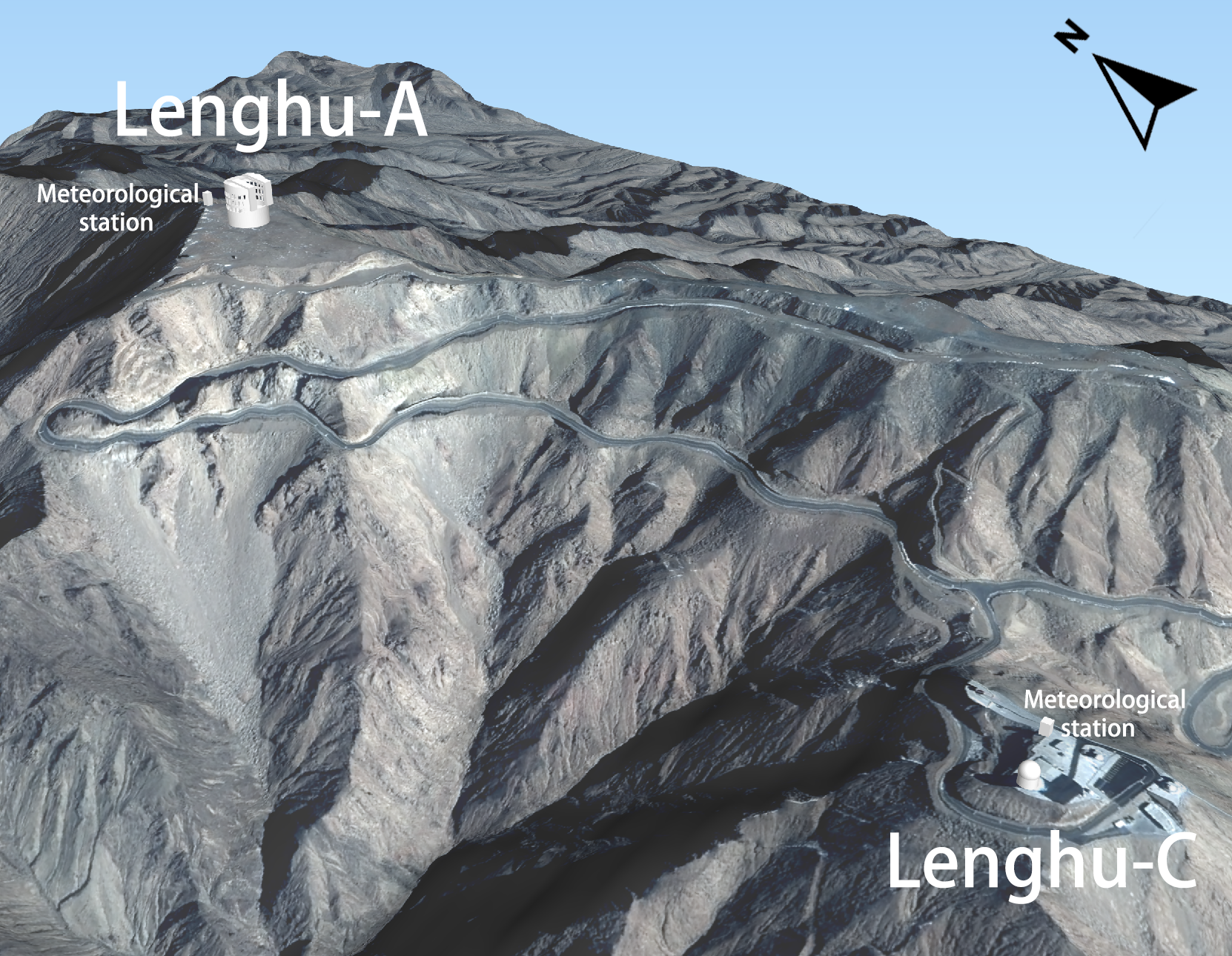}
  \caption{Three-dimensional view of Lenghu-A and Lenghu-C. Lenghu-C was completed in 2023, and the site levelling for Lenghu-A was finished in the same year. Each meteorological station is close to the northwest slope of the respective platform.\label{fig3}}
\end{figure}

The meteorological stations at Lenghu-A and Lenghu-C (Figure \ref{fig2}) were installed in 2024 and 2018, respectively. Each meteorological station is erected on a 10 m tower located close to the steep slope northwest of the respective platform. The horizontal distance between the two towers is approximately 1.1 km. Figure \ref{fig3} shows the three-dimensional view of the two platforms, including the simplified models of meteorological stations and telescopes. The satellite image was produced from the Multi-Spectral Scanner(MSS) and Panchromatic (PAN) data of Jilin-1. 

\begin{figure}[ht!]
  \centering
  \includegraphics[width=10cm]{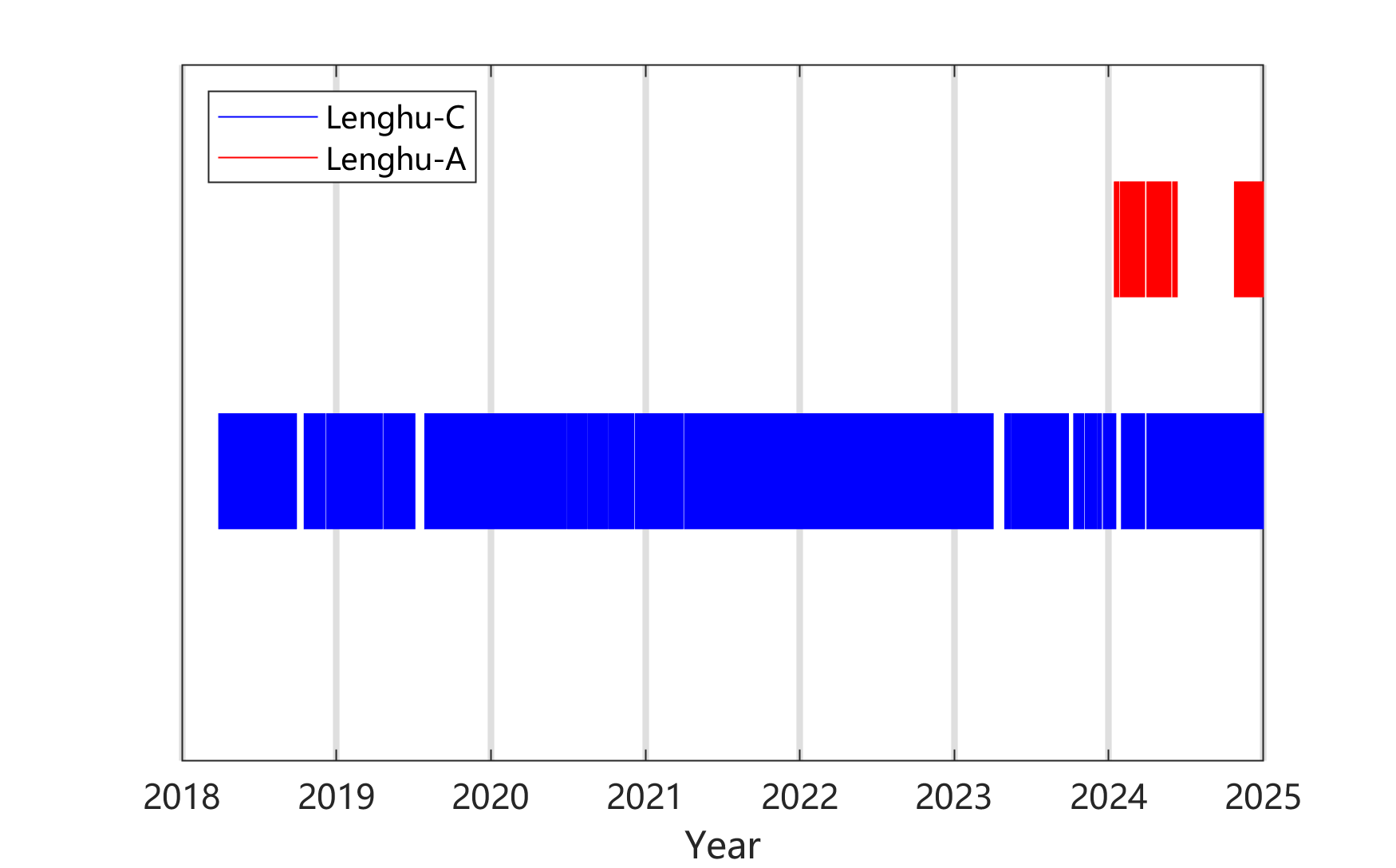}
  \caption{Time coverage of meteorological data from Lenghu-A (short term) and Lenghu-C (long term), as of December 2024.\label{fig4}}
\end{figure}

Figure \ref{fig4} shows the time coverage of meteorological data from Lenghu-A starting in January 2024 and Lenghu-C starting in March 2018. The blank entries indicate missing data caused by power outages, equipment failures, and other miscellaneous mishaps. 

The meteorological station at Lenghu-C was relocated to its current location in February 2023 \citep{10.1088/1674-4527/adcefe}, the blocking or channeling effects of the surrounding terrain may lead to changes in observed wind directions \citep{2022RAA....22d5002L}.Therefore, in this paper, the wind direction analysis for Lenghu-C is based on data after February 2023, other analyses utilize the entire data set. All data has been preprocessed, including timestamp deduplication, zero data removal (consecutive zero values exceeding 10 entries), and hourly outlier rejection based on 3-sigma criteria.

\subsection{Circular statistics method for wind direction data} \label{sec:2.3}
The wind direction data is expressed in degrees from 0° to 360°, its value range forms a closed circle (e.g., 0° = 360°). Traditional statistical methods, such as arithmetic mean, cannot deal with this periodicity. For example, if the wind direction is 350° and 10°, their arithmetic mean is 180°, but the actual average direction should be 0°. Therefore, when processing wind direction data, the circular statistics method must be used. All circular statistics (circular means, correlations \citep{10.1093/biomet/70.2.327}) were calculated with CircStat toolbox \citep{JSSv031i10}. Besides, the Pearson correlation coefficient was also calculated for time-series wind direction data comparison of Lenghu-A and Lenghu-C.

\section{Results and Discussion} \label{sec:3}
The statistical data of temperature and wind speed is shown in Table \ref{tab:1}. 
According to data attained in 2024, Lenghu-A experienced lower temperatures and higher wind speeds, which may be attributed to its higher elevation. Consequently, telescopes and enclosures installed at Lenghu-A may require specific adjustments to address these environmental challenges.

\begin{table}[ht!]
\centering
\caption{Temperature and wind speed data of Lenghu-A and Lenghu-C.}
\label{tab:1}
\begin{tabular}{llll}
\hline
Temperature (°C)     & Lenghu-A (2024) & Lenghu-C (2024) & Lenghu-C (2018-2024) \\ \hline
Max             & 18.7   & 20.2  & 29.1     \\
Min             & -25.8  & -23.2  & -27.3    \\
Average         & -8.2   & -2.0  & -1.7     \\
Median          & -8.9   & -1.6  & -1.6     \\
Nighttime median    & -10.6  & -2.4  & -2.5     \\
Max daily range & 17.7   & 13.4  & 19.3     \\ \hline
Wind speed (m/s)      & Lenghu-A (2024) & Lenghu-C (2024) & Lenghu-C (2018-2024)\\ \hline
Max             & 19.4   & 23.0   & 41.7     \\
Average         & 3.8    & 3.2  & 4.5      \\
Median          & 3.4    & 2.5  & 3.7      \\
Nighttime median    & 3.9    & 2.6  & 4.0      \\ \hline
\end{tabular}
\end{table}

\subsection{Temperature} \label{sec:3.1}
\subsubsection{Hourly temperature}\label{sec:3.1.1}
The thermal management system of large telescope enclosures requires precise cooling to achieve the target temperature equivalent to midnight ambient temperature, facilitating rapid thermal equilibrium between the internal and external environments upon slits opening. Additionally, the cooling load calculation for the air-conditioning system requires continuous hourly temperature profiles to inform equipment selection and operational strategies of the enclosure cooling infrastructure.

\begin{figure}[ht!]
  \centering
  \begin{subfigure}{0.4\textwidth}
    \includegraphics[width=\linewidth]{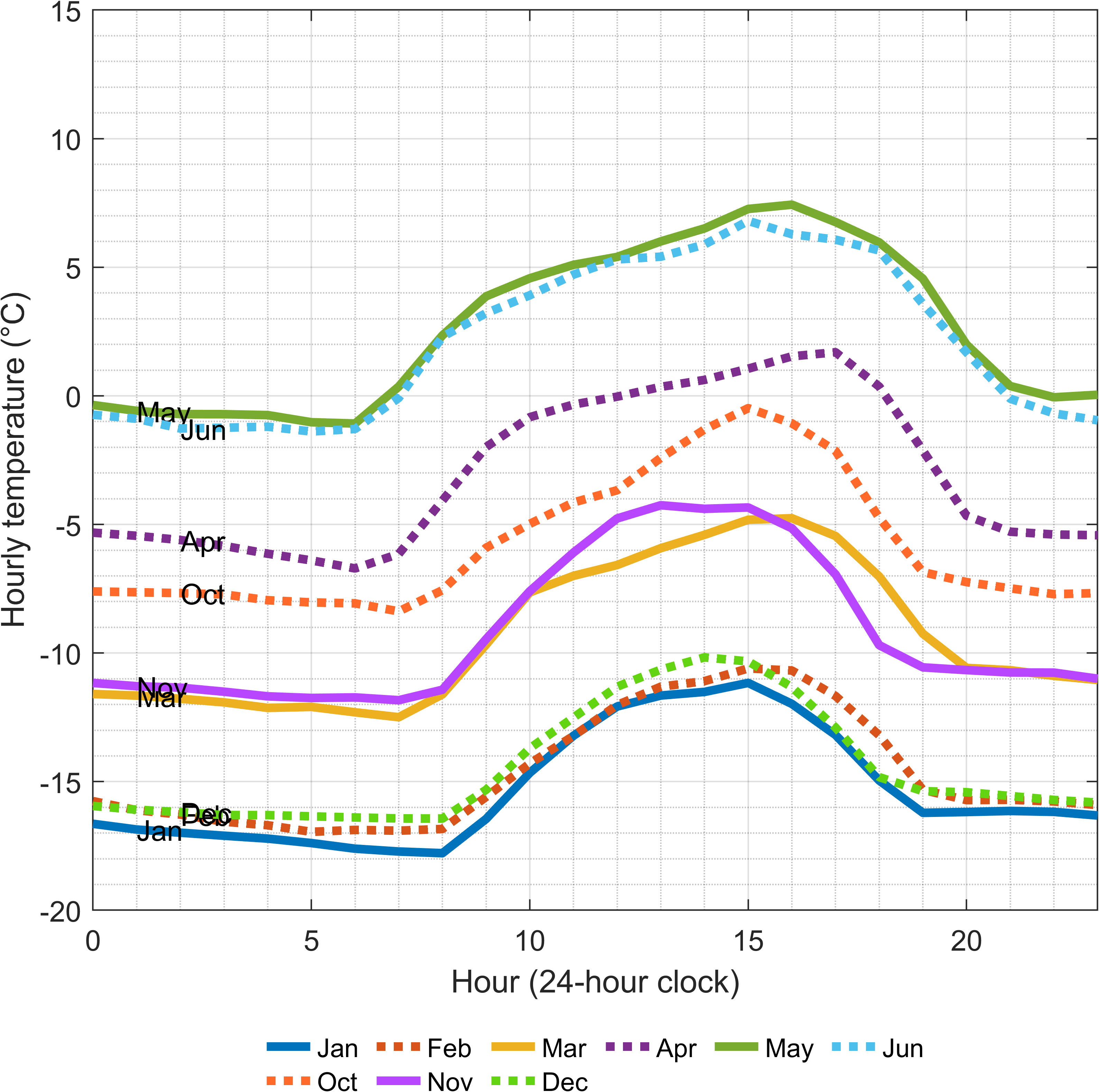}
  \end{subfigure}
  \hspace{1em}
  \begin{subfigure}{0.4\textwidth}
    \includegraphics[width=\linewidth]{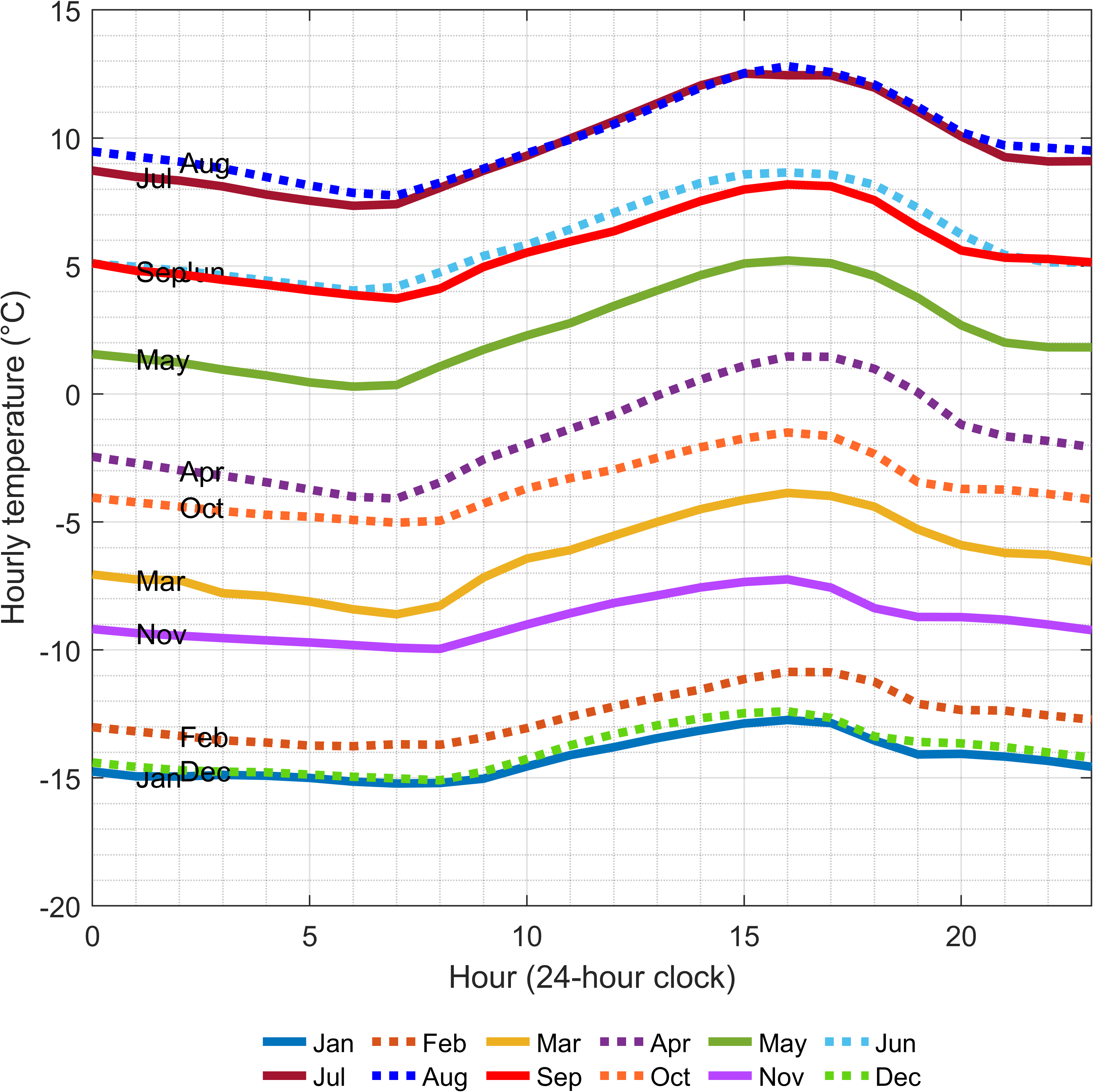}
  \end{subfigure}
  \caption{Hourly average temperature of Lenghu-A (left) and Lenghu-C (right). The data for Lenghu-A lacks records from July to September.}
  \label{fig5}
\end{figure}

\begin{figure}[ht!]
  \centering
  \includegraphics[width=10cm]{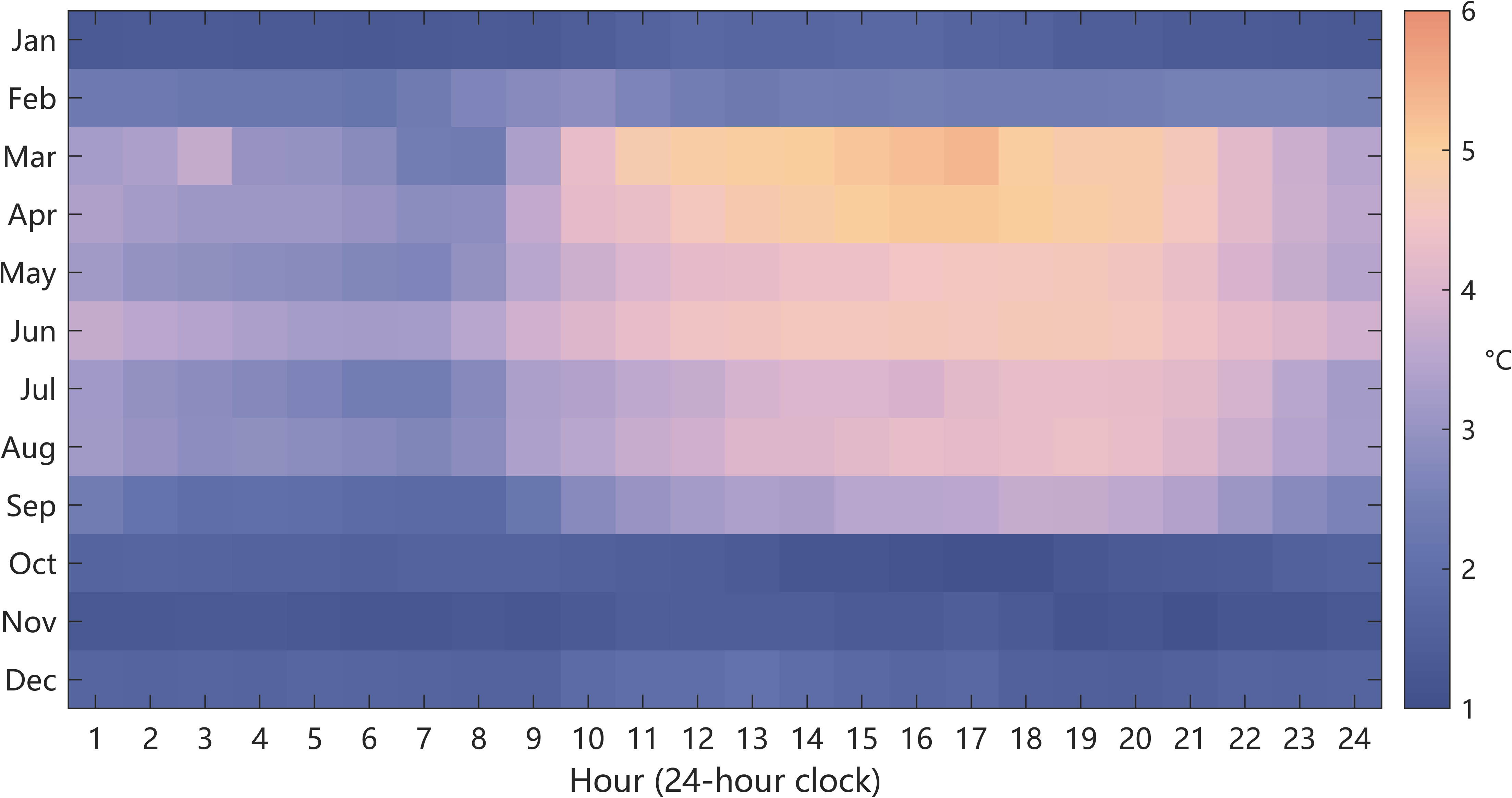}
  \caption{The interannual standard deviation of hourly average temperature of Lenghu-C. \label{fig6}}
\end{figure}

Figure \ref{fig5} presents the hourly average temperature for each month, with annual averaging applied to the data of Lenghu-C. Lenghu-A has larger fluctuations (max. 8.5°C in May), while Lenghu-C showed more minor variations (max. 5.5°C in April). However, from March to September, the hourly temperature fluctuations during daytime exhibit significant interannual variability, with a maximum variation of 6°C (Figure \ref{fig6}).

\begin{figure}[ht!]
  \centering
  \begin{subfigure}{0.4\textwidth}
    \includegraphics[width=\linewidth]{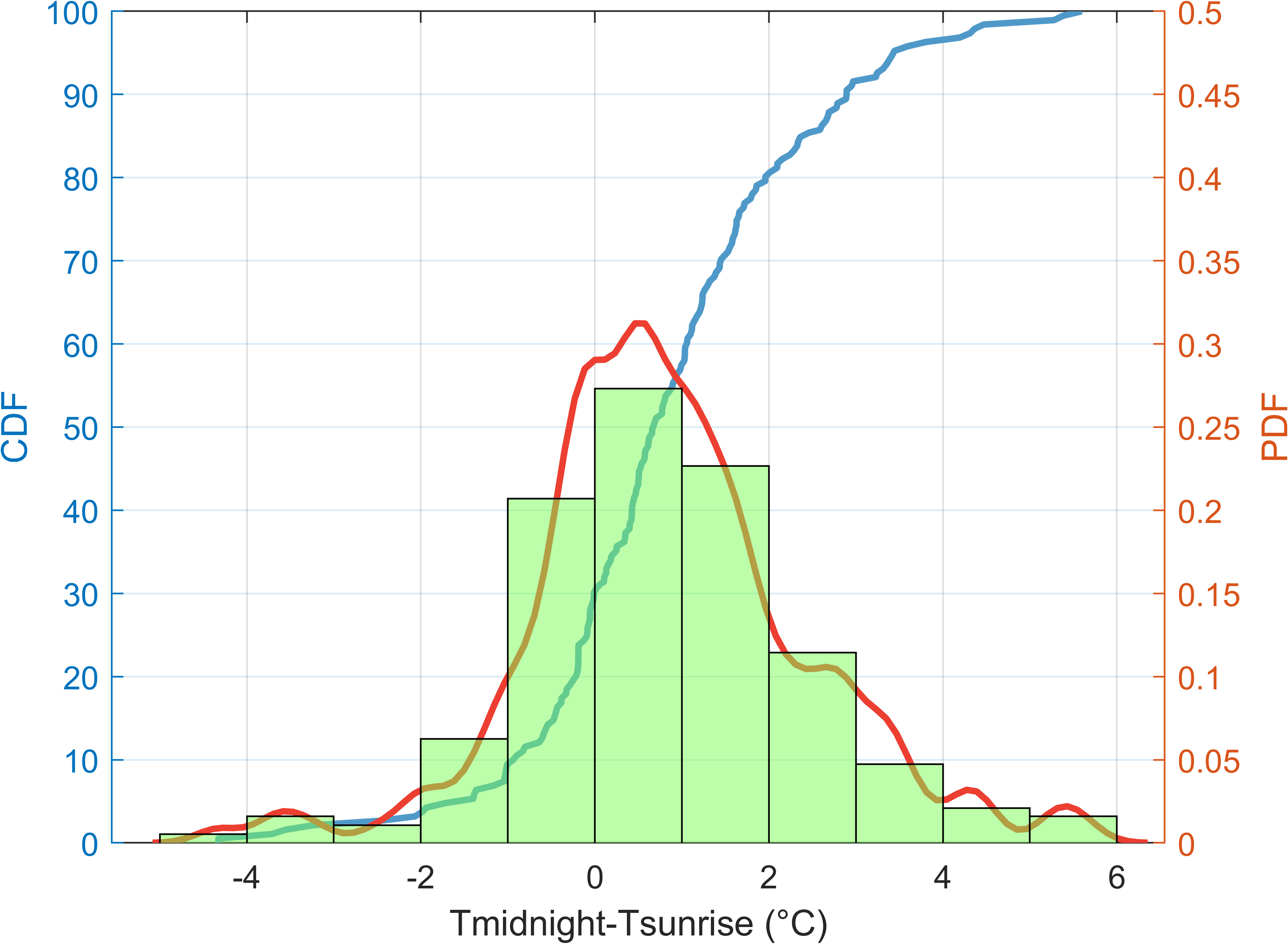}
  \end{subfigure}
  \hspace{1em}
  \begin{subfigure}{0.4\textwidth}
    \includegraphics[width=\linewidth]{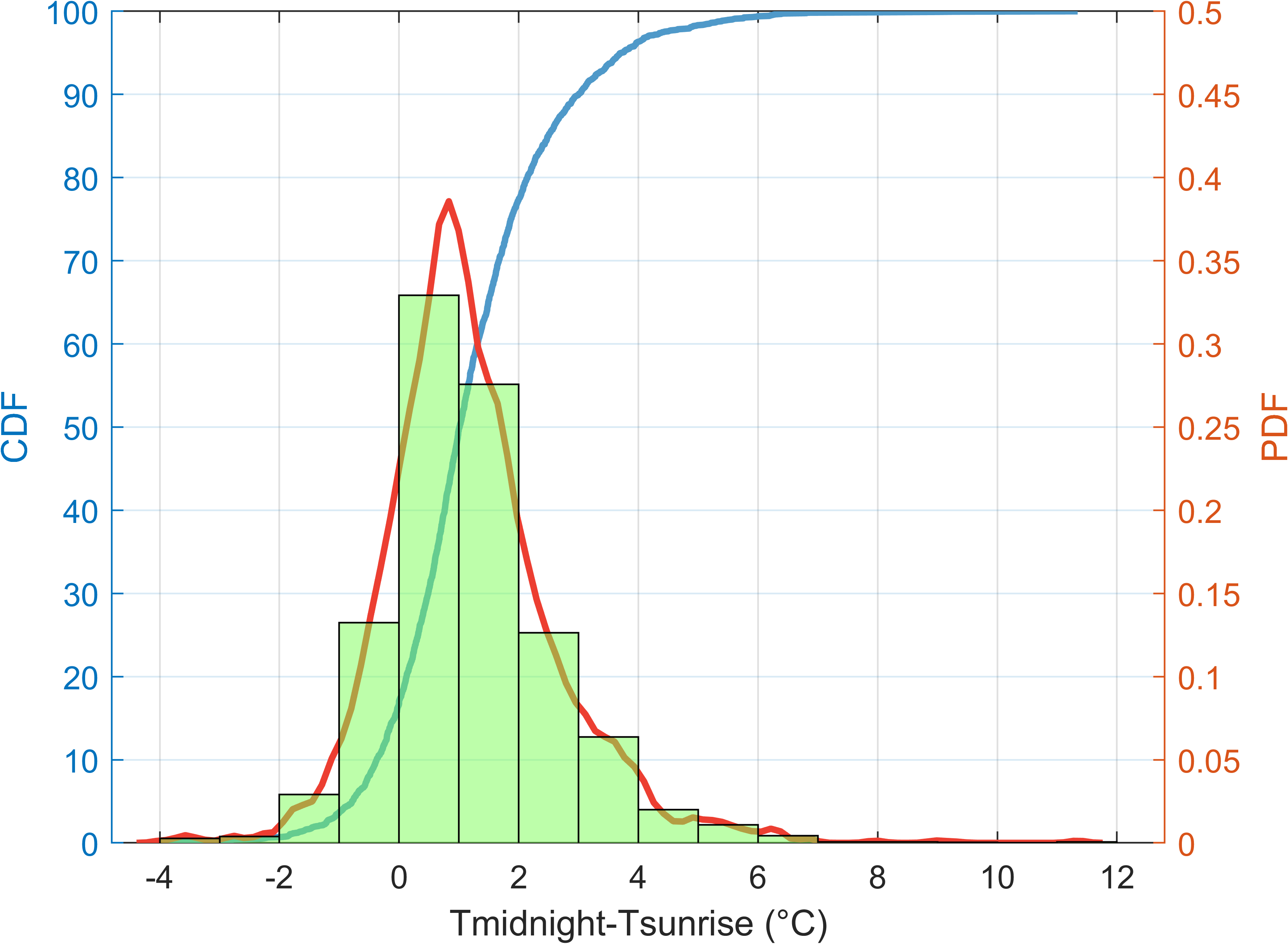}
  \end{subfigure}
  \caption{ Distribution of the temperature difference between midnight and sunrise for Lenghu-A (left) and Lenghu-C (right).} 
  \label{fig7}
\end{figure}

Figure \ref{fig7} shows the distribution of temperature differences between sunrise (\textit{T$_{\textrm{sunrise}}$}) and midnight (\textit{T$_{\textrm{midnight}}$}), with the Cumulative Distribution Function (CDF) and Probability Density Function (PDF) curves, based on data in 2024. The enclosure will be closed at sunrise, the air inside will begin warming due to solar radiation. It is hoped that the air cooling system could rapidly bring the overall air temperature close to the midnight temperature while minimizing the vertical gradients within the enclosure. For both Lenghu-A and Lenghu-C, the temperature differences between midnight and sunrise are expected to be ±2℃ within approximately 80\% of the time. Computational Fluid Dynamics (CFD) analysis should be conducted during the enclosure design to address these cases.

To optimize computational resources and specify the time periods for CFD simulations of cooling capacity, a seasonal analysis of hourly temperature variations was conducted to calculate the maximum accumulation of temperature differences between hourly values from sunrise to sunset (\textit{T}) and midnight, thereby identifying the month with the highest cooling capacity demand (Table \ref{tab:2}).


\begin{table}[]
\centering
\caption{The maximum accumulation of temperature differences for each season at Lenghu-A and Lenghu-C. The months presented in the table correspond to the maximum accumulation of \textit{T-T$_{\textrm{midnight}}$}.}
\label{tab:2}
\begin{tabular}{lcccc}
\hline
\hline
\multirow{2}{*}{Unit: °C} & \multicolumn{4}{c}{Lenghu-A(Lenghu-C)}          \\ \cline{2-5} 
                        & Spring & Summer & Autumn & Winter \\ \hline
Month                   & May(Apr)      & Jun(Jun)      & Nov(Sep)     & Dec(Feb)    \\
sum(\it{T-T$_{\textrm{midnight}}$})       & 70.9(23.0)   & 68.9(23.8)   & 48.6(15.8)   & 35.5(10.8) \\
\it{T$_{\textrm{midnight}}$}          & -0.4(-2.5)   & -0.7(5.1)   & -11.2(5.1)  & -15.9(-13.0)  \\
\it{T$_{\textrm{sunrise}}$}           & -1.0(-4.0)   & -1.4(4.2)   & -11.7(3.9)  & -16.4(-13.7) \\
\it{T$_{\textrm{sunset}}$}           & 2.0(-1.2)    & 1.7(6.2)    & -10.7(6.5)  & -15.4(-12.1)  \\
\it{T$_{\textrm{max}}$}              & 7.4(1.5)    & 6.8(8.7)    & -4.3(8.2)   & -10.2(-10.9)   \\
\it{T$_{\textrm{min}}$}              & -1.1(-4.1)   & -1.4(4.0)   & -11.8(3.7)  & -16.4(-13.8)  \\ \hline
\end{tabular}
\end{table}

By integrating hourly temperature, solar radiation, internal heat sources (e.g., Motors and instruments), and the thermal properties of the building envelope (including conductivity and heat transfer coefficients), a preliminary cooling load estimation for the enclosure can be conducted using established engineering methods \citep{GB2012,ASHRAE2017,Omar_2022}. For a detailed analysis, CFD can be employed to conduct hourly transient simulations of heat distribution within the enclosure, enabling accurate determination of the total cooling load. The simulation periods in May and June are appropriate for Lenghu-A and Lenghu-C.

\subsubsection{Daily temperature variation}\label{sec:3.1.2}
\begin{figure}[ht!]
  \centering
  \includegraphics[width=12cm]{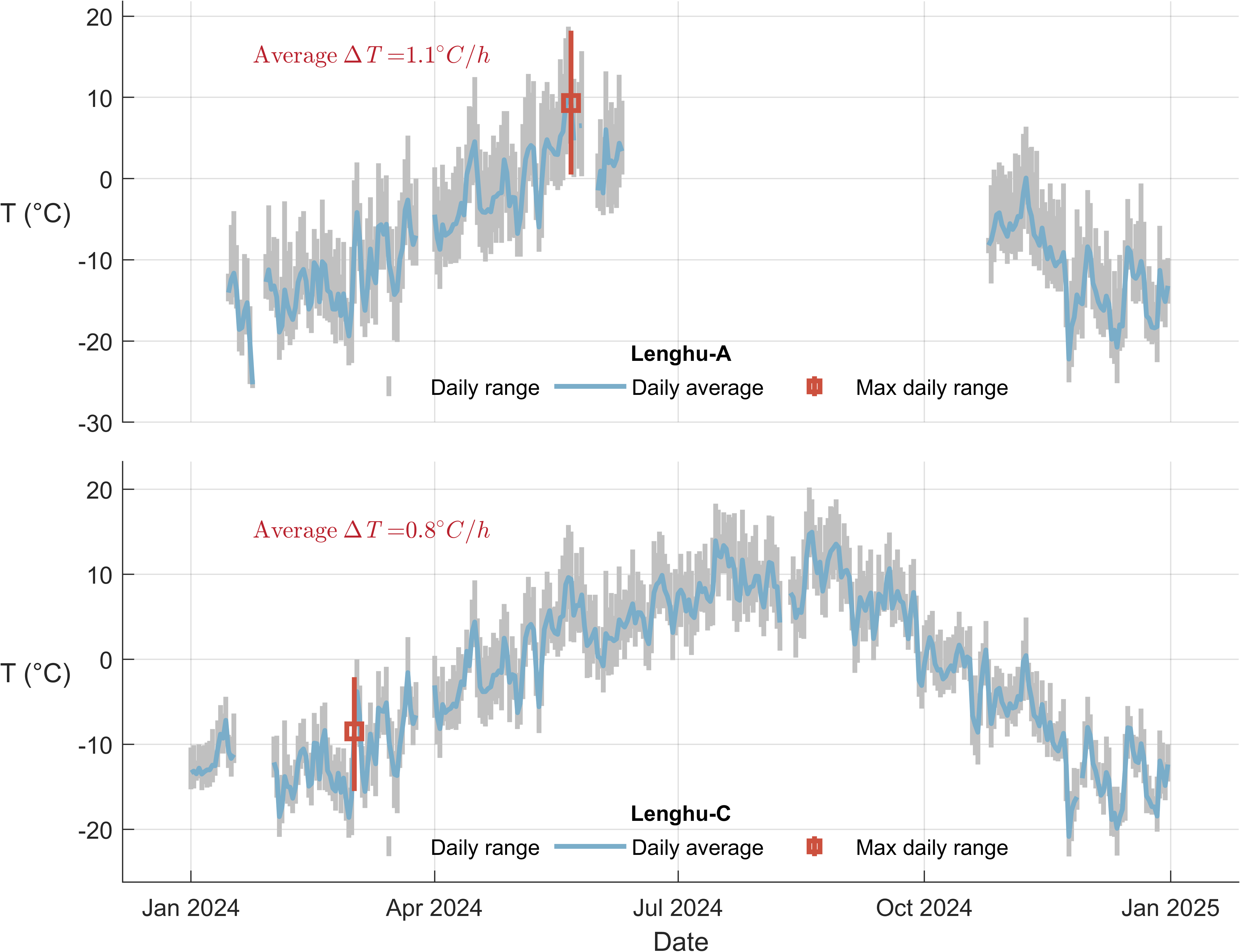}
  \caption{Daily temperature variation of Lenghu-A (top) and Lenghu-C (below) in 2024. $\Delta T$ is the annual average temperature variation rate. The grey and blue lines present the daily temperature range and the daily average temperature, respectively. \label{fig8}}
\end{figure}

The daily temperature ranges are visualized on a graph to illustrate the variation over a year. As shown in Figure \ref{fig8}, the daily temperature range and the daily average temperature are presented with grey and blue lines, respectively. Temperature at the Lenghu site remains relatively stable, with an annual average temperature variation rate ($\Delta T$) of approximately 1°C per hour. The average daily temperature range was 6.7°C, based on temperature data from Lenghu-C, whereas Lenghu-A's data, which lacked records from July to September, exhibited a higher daily temperature range of 10.2°C.

\subsection{Wind speed} \label{sec:3.2}
\subsubsection{Wind speed distribution} \label{sec:3.2.1}
The statistics of wind speed distribution contain the entire data set. Figure 9 presents the wind speed histograms, with the 75th and 50th percentiles explicitly indicated. The wind speed distributions of the two platforms are generally consistent, with 75th percentile of 5.3 m/s and 6.1 m/s, and 50th percentile of 3.4 m/s and 3.7 m/s, respectively. For the CFD analysis of wind and temperature fields of telescope and enclosure under normal conditions, the nighttime median wind speed, i.e., 4 m/s, should be employed. 

\begin{figure}[ht!]
  \centering
  \begin{subfigure}{0.4\textwidth}
    \includegraphics[width=\linewidth]{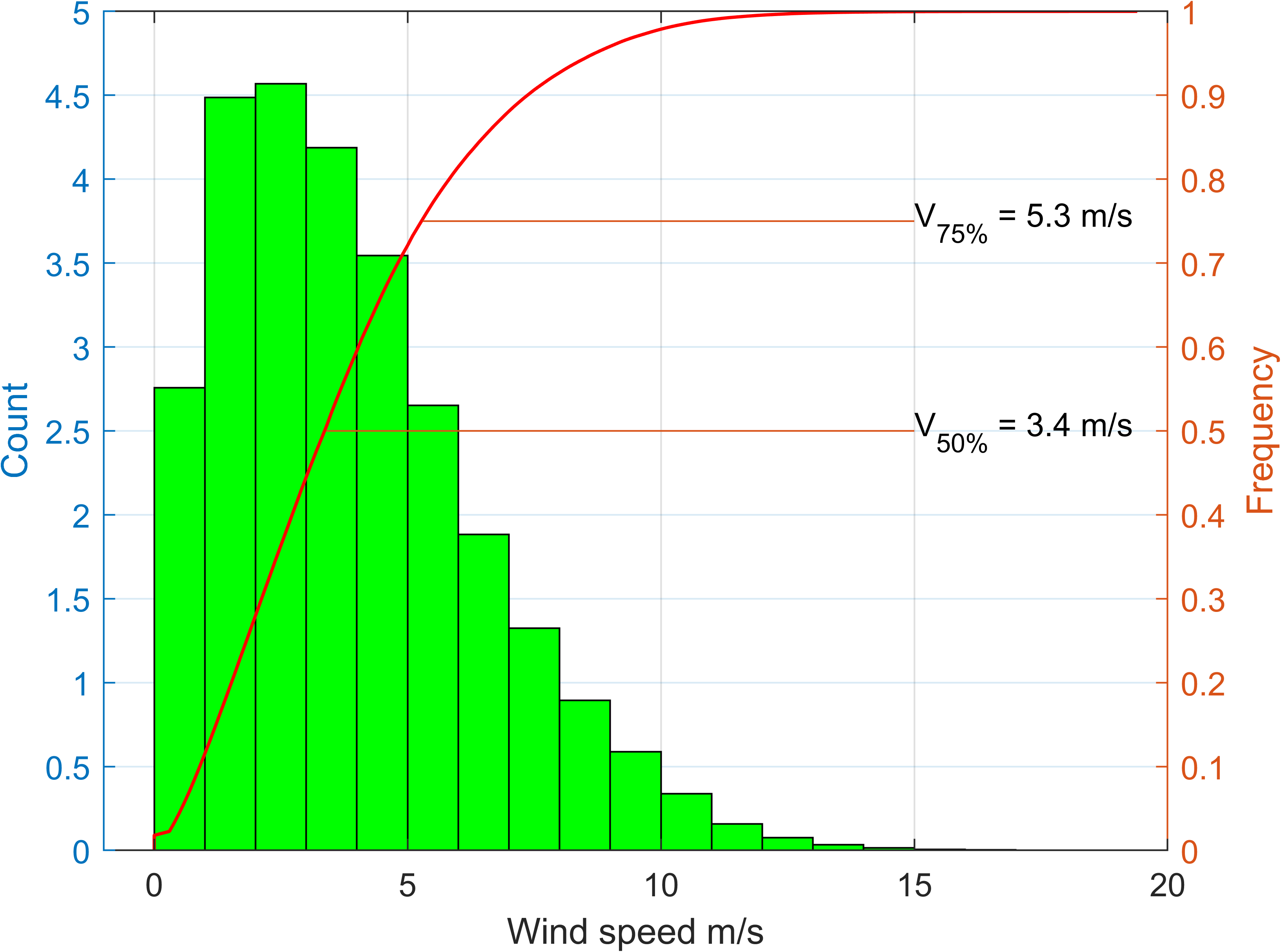}
  \end{subfigure}
  \hspace{1em}
  \begin{subfigure}{0.4\textwidth}
    \includegraphics[width=\linewidth]{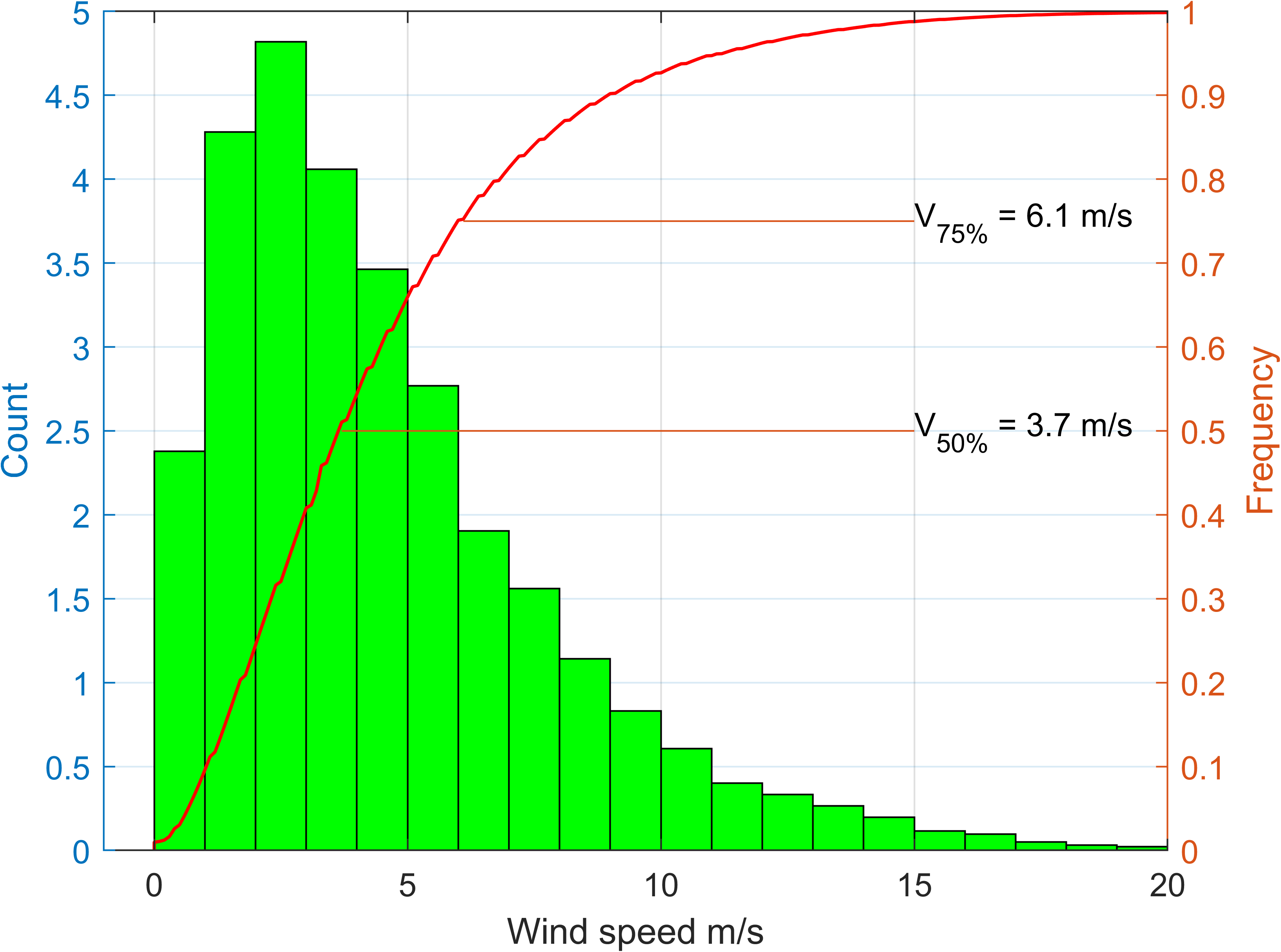}
  \end{subfigure}
  \caption{Wind speed distributions of Lenghu-A (left) and Lenghu-C (right), with red horizontal lines indicating the 75th and 50th percentiles. The $x$-axis was limited to 0-20 m/s.} 
  \label{fig9}
\end{figure}

\begin{figure}[ht!]
  \centering
  \includegraphics[width=12cm]{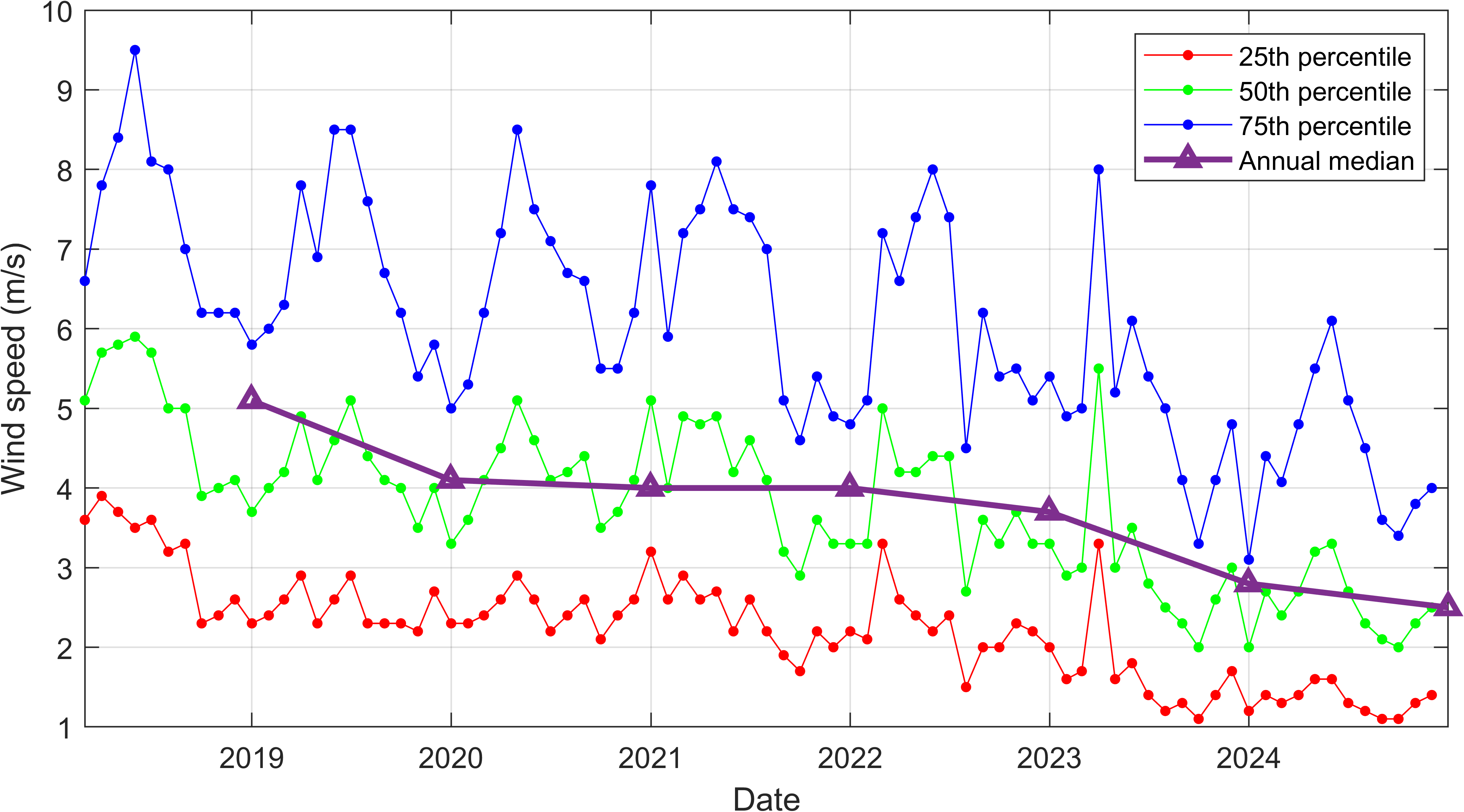}
  \caption{Monthly average wind speed of Lenghu-C during 2018-2024, with 25th, 50th and 75th percentiles. The annual median values are marked with triangle.\label{fig10}}
\end{figure}

The wind speed has gradually decreased in recent years (Figure \ref{fig10}), with an annual decrease rate of approximately 0.37 m/s in the median wind speed, causing earlier higher wind data to elevate the multi-year median compared to shorter-term measurements at Lenghu-A. Furthermore, throughout the measurement period, May and June record the highest monthly average wind speed.

The maximum wind speed recorded at Lenghu-A was 19.4 m/s (in 2024, see Table \ref{tab:1}). In contrast, Lenghu-C exhibited a significantly higher maximum of 41.7 m/s, which occurred on September 28, 2018. This disparity highlights the critical importance of sufficient data duration for capturing extreme wind events, as short-term measurements may underestimate potential maximum wind speeds. The statistical parameters demonstrate that while central tendency measures (e.g., 50th percentiles) show similarity between two platforms, extreme values differ substantially due to temporal sampling limitations. Such results emphasize the necessity of long-term wind monitoring for accurate enclosure structural safety evaluations.

\subsubsection{Extreme wind speed} \label{sec:3.2.2}
Given the limited availability of long-term data, the extreme wind speed of Lenghu-A was estimated by scaling the maximum wind speed of Lenghu-C (41.7 m/s) with the ratio of their average wind speeds (2023-2024), i.e., $\textrm{41.7} \times (\textrm{3.8/3.2})\approx \textrm{49.5}$ m/s. In addition, the power law model is used to describe the wind profile with height $V(z)$,
\begin{equation}\label{equ:1}
  V(z)=V_{ref}(\frac{z}{z_{ref}})^{\alpha}
\end{equation}
Where $V_{ref}$ is the known wind speed at a reference height,$z$ is the current height, $z_{ref}$ is the reference height, $\alpha$ is the empirical exponent that varies depending on the stability of the atmosphere. For the telescopes at Lenghu site, a terrestrial wind profile with exponent $\alpha = \textrm{0.16}$ is assumed \citep{2016SPIE.9911E..14L}. For a telescope at Lenghu-A, assuming the height of enclosure is approximately 60 m, $V(\textrm{60}) = \textrm{65.9} $ m/s. Therefore, the wind resistance capability of enclosure structures under survival condition of the enclosure must exceed $V(\textrm{60})$.  
Other methods to determine the wind speed for survival conditions could be also adopted, such as the 50-year return period extreme wind speed \citep{2016SPIE.9906E..10T,gmt2021}, or referring to relevant national standards in the construction \citep{GB50009} and wind energy field \citep{NB2018} . 

\subsubsection{Hourly wind speed} \label{sec:3.2.3}
\begin{figure}
  \centering
  \includegraphics[width=12cm]{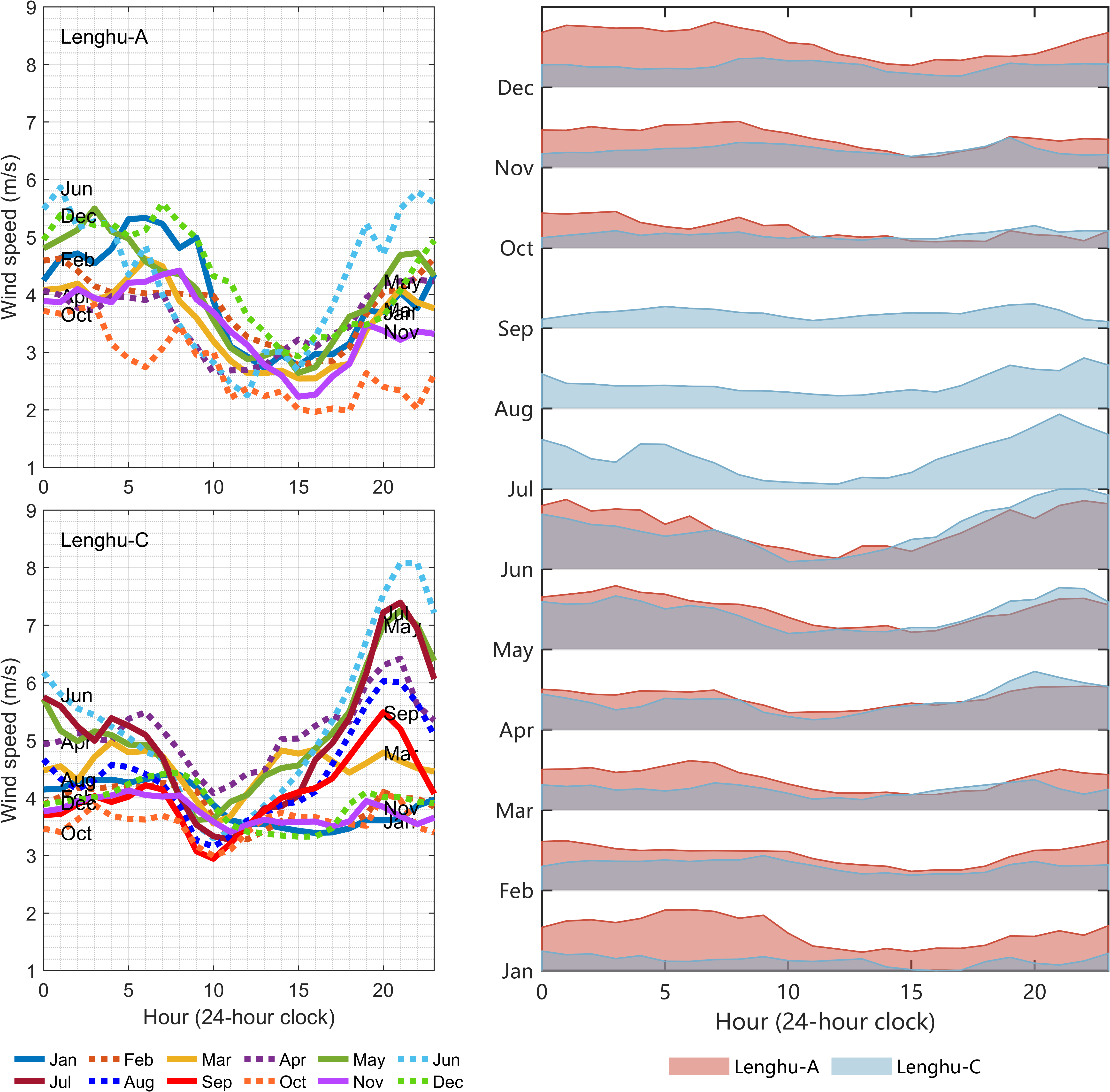}
  \caption{Hourly average wind speed of Lenghu-A (upper left) and Lenghu-C (lower left). The right panel presents two wind speed curves for each month in 2024, plotted on the same graph to facilitate comparative analysis of their monthly variations. \label{fig11}}
\end{figure}

\begin{figure}
  \centering
  \includegraphics[width=10cm]{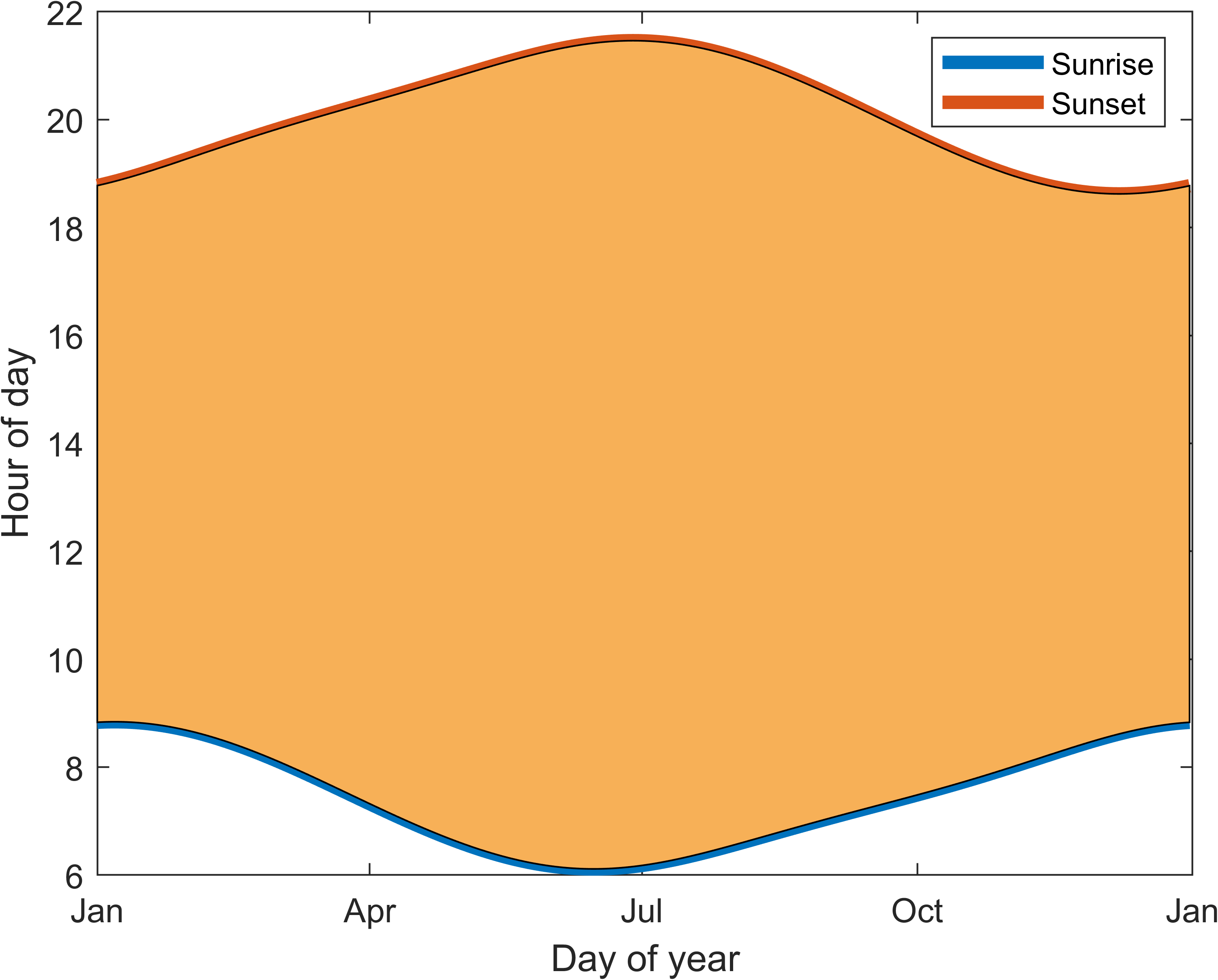}
  \caption{Sunrise and sunset times at Lenghu site in 2024. The coordinate used for calculation is 38.6°N, 93.9°E, 4300 m. \label{fig12}}
\end{figure}

Figure \ref{fig11} shows the hourly average wind speed for each month. On the right panel, wind speed curves of Lenghu-A and Lenghu-C are plotted on the same graph to facilitate comparative analysis of their monthly variations (in 2024). Overall, the nighttime wind speeds are generally higher than daytime values. Hourly wind speed variations follow a characteristic pattern from sunrise to sunset (Sunrise and sunset times at Lenghu site in 2024 is shown in Figure \ref{fig12}), which is an initial decrease followed by a gradual increase. The minimum wind speed occurs around 15:00 at Lenghu-A, while the corresponding transitional phase appears near 10:00 at Lenghu-C. This phenomenon is particularly pronounced during the April-September period.

\subsection{Wind direction} \label{sec:3.3}
\subsubsection{Wind direction distribution} \label{sec:3.3.1}
\begin{figure}
  \centering
  \includegraphics[width=12cm]{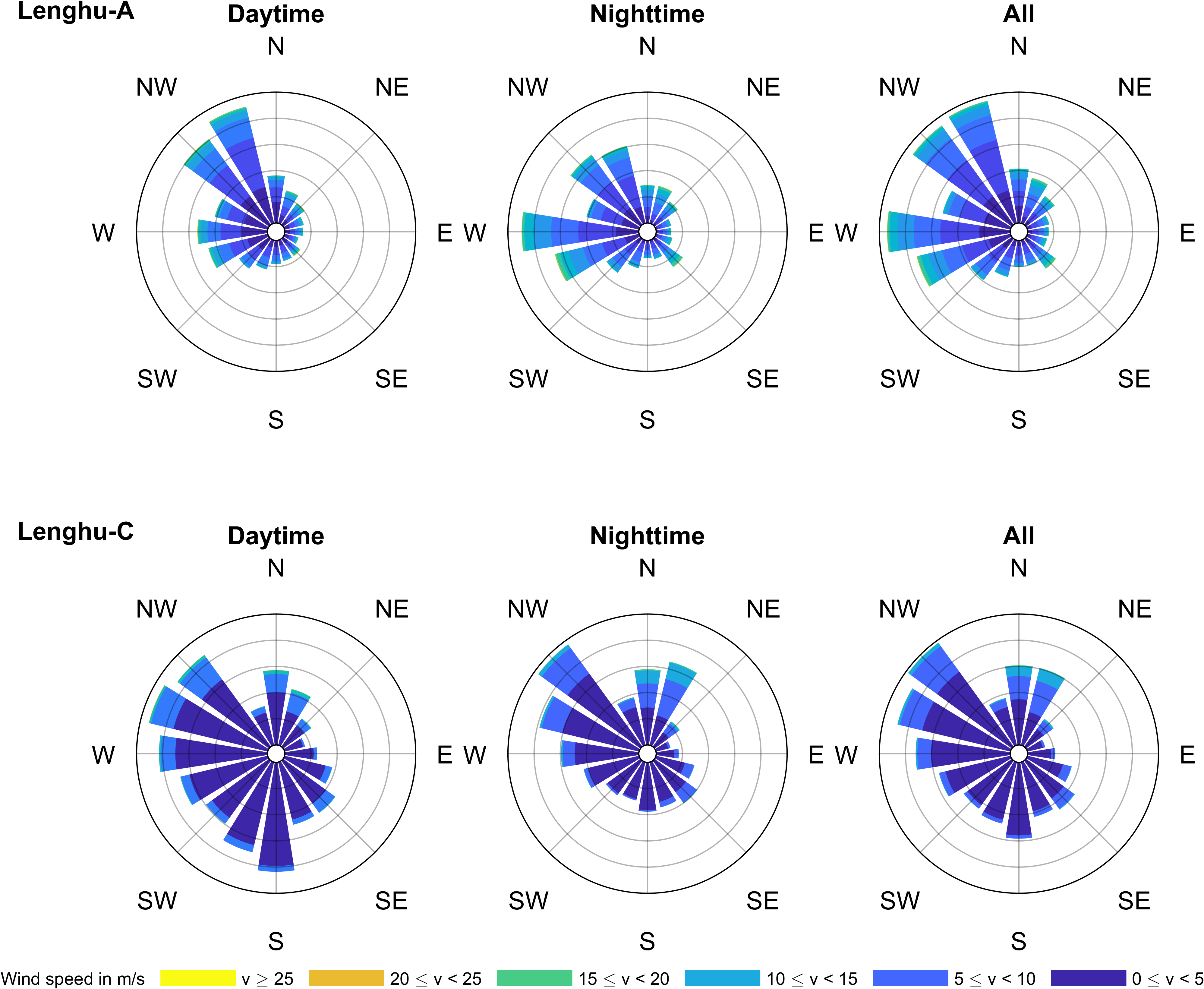}
  \caption{Wind direction distributions of Lenghu-A and Lenghu-C, illustrating the wind rose diagrams of daytime periods, nighttime periods and all day.} 
  \label{fig13}
\end{figure}

According to the wind rose diagrams in right column of Figure \ref{fig13}, the prevailing wind direction at Lenghu-A and Lenghu-C is NNW (north-northwest) and NW (northwest), respectively. During the initial design phase, CFD analysis should be conducted to evaluate the effects of terrain and multiple telescopes based on the prevailing wind direction, thereby determining the optimal installation locations. 

\begin{figure}
  \centering
  \includegraphics[width=12cm]{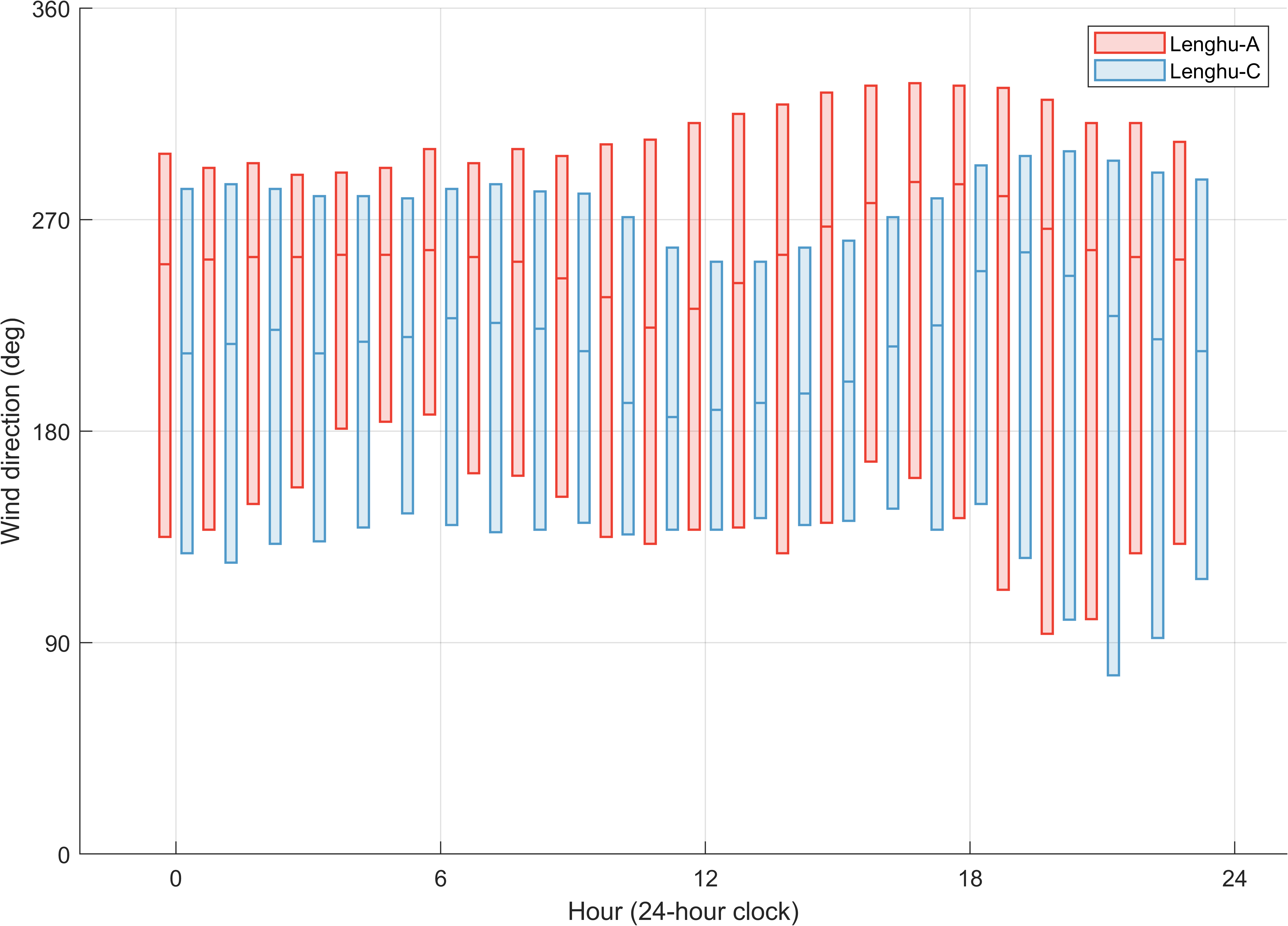}
  \caption{Hourly wind direction of Lenghu-A (red box) and Lenghu-C (blue box) in 2024. The line inside of each box is the median. The top and bottom edges of each box are the 75th and 25th percentiles, respectively. \label{fig14}}
\end{figure}

Hourly wind direction analysis (circular mean) was performed to investigate its variation patterns, thereby providing guidance for ventilation strategy research during nighttime observation. Figure \ref{fig14} presents a box chart, displaying the median, the lower and upper quartiles (25th and 75th percentiles). The whiskers, minimum and maximum values are omitted because the minimum and maximum wind directions are always 0° and 360°, which would result in a cluttered visualization. The hourly wind direction at Lenghu-A differed significantly from that at Lenghu-C, although their median variation trends were similar. The median wind direction at Lenghu-A was approximately 40 degrees more northerly than that at Lenghu-C. As shown in Figure \ref{fig13} and Figure \ref{fig14}, Lenghu-A predominantly experienced NNW and NW winds during daytime, while west wind prevailed at night. In contrast, the daytime and nighttime wind directions at Lenghu-C were dominated by WNW (west-northwest) and NW winds, respectively, with a higher proportion of south wind measured during daytime. 

\subsubsection{Wind direction correlation} \label{sec:3.3.2}
\begin{figure}
     \centering
     \begin{subfigure}{\linewidth}
       \centering
       \includegraphics[width=0.8\linewidth]{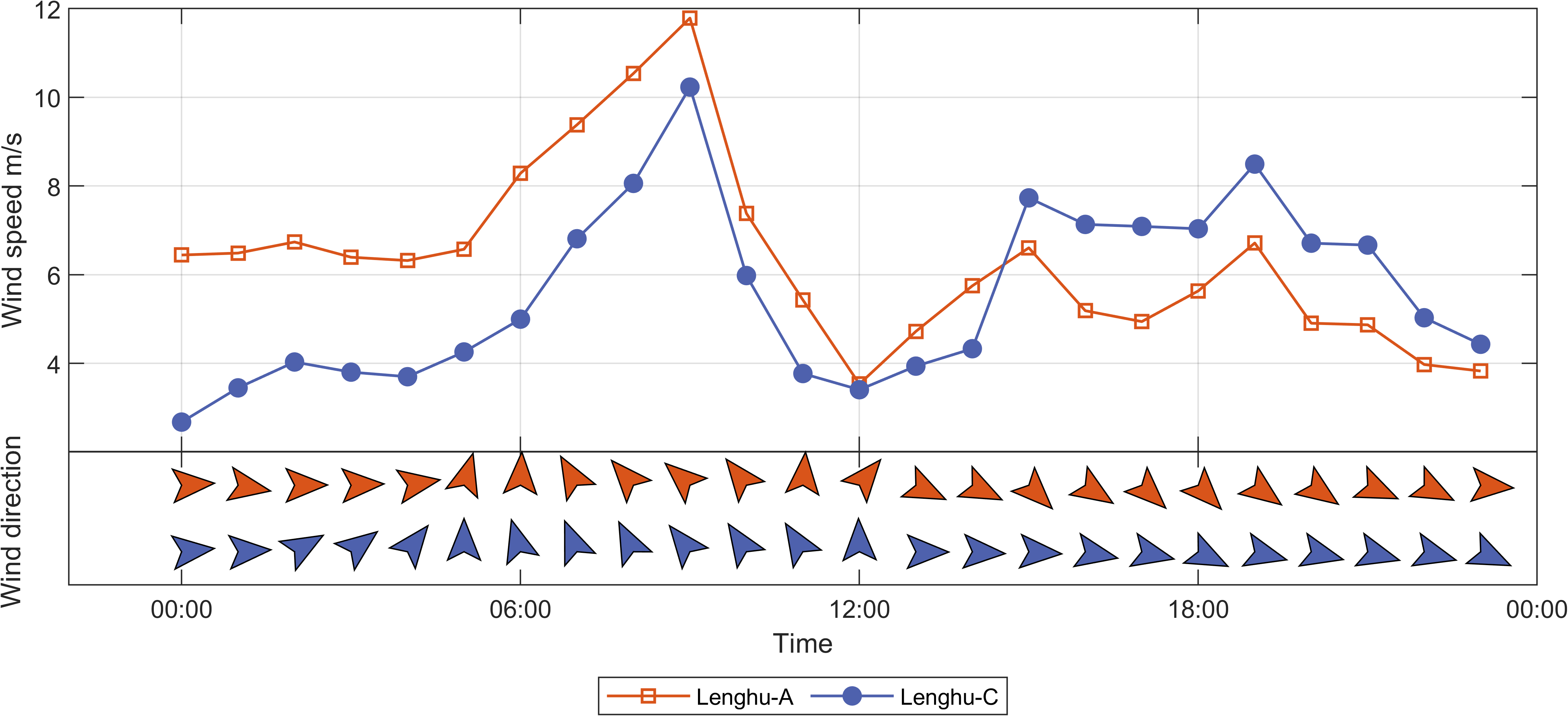}
       \label{fig:a}
     \end{subfigure}
     \begin{subfigure}{\linewidth}
       \centering
       \includegraphics[width=0.8\linewidth]{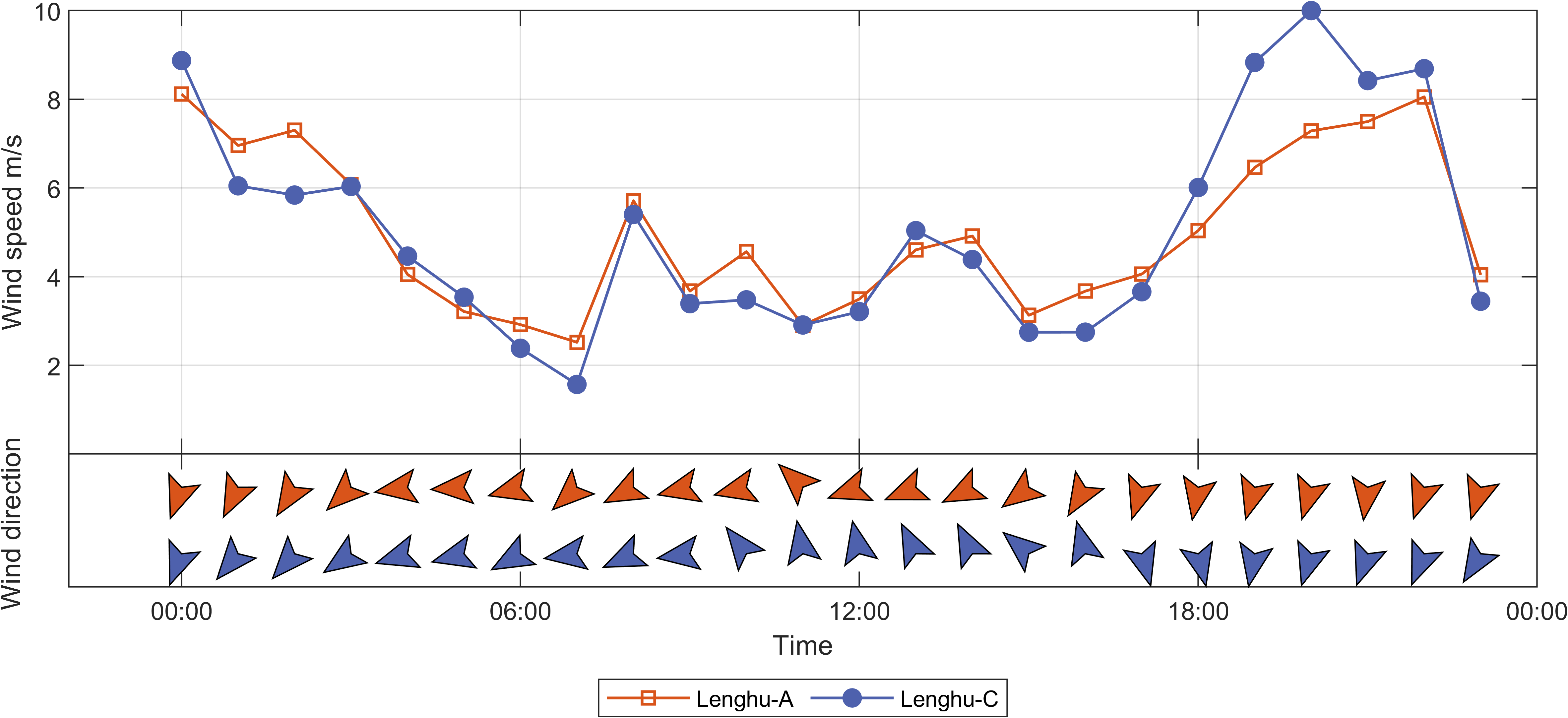}
       \label{fig:b}
     \end{subfigure}
     \caption{Time-series variation of wind speed and wind direction on February 17, 2024 (top) and May 01, 2024 (below). The value for each hour was calculated by hourly average. Specifically, the wind speed data were averaged using arithmetic mean, while the wind direction data were using the circular mean method to process the 0°–360° discontinuity.}
     \label{fig15}
\end{figure}

Wind direction correlation was analyzed using hourly average time-series data with identical time stamps in 2024, encompassing 189 days of complete 24-hour continuous records. Figure \ref{fig15} presents daily variations in wind speed and direction on two representative days, illustrating the fluctuations and trends between Lenghu-A and Lenghu-C. The wind speed and direction between the two platforms exhibited overall consistency, though discrepancies were identified during specific time intervals. 

The correlation coefficients over the entire year were calculated from the correlation coefficient for each day. The median Pearson correlation coefficient and the circular correlation coefficient are 0.70 and 0.68, respectively, indicating a strong correlation between wind directions at Lenghu-A and Lenghu-C. Meanwhile, the median Pearson correlation coefficient of temperature and wind speed are 0.65 and 0.88, respectively.

\section{Conclusion} \label{sec:4}
In this paper, new statistics of temperature and wind characteristics of the Lenghu site (Lenghu-A and Lenghu C) are presented based on the latest measurements. These statistics are especially important for the thermal and structural design of the telescope and enclosure, as well as the design of the ventilation and cooling systems of the enclosure. These results could be helpful for telescope projects that are planning to be located at the Lenghu site. If anyone is interested in obtaining these statistical results in data format, please contact the authors of this paper.

\bibliographystyle{raa}
\bibliography{raa-lenghu}


\end{document}